\documentclass[conference,a4paper]{IEEEtran}

\usepackage{graphicx}
\usepackage{textcomp}
\usepackage{array}
\usepackage{multirow}
\usepackage{underscore}
\usepackage{geometry}
\newcolumntype{P}[1]{>{\centering\arraybackslash}p{#1}}

\newenvironment{conditions}[1][where:]
  {#1 \begin{tabular}[t]{>{$}l<{$} @{${}={}$} l}}
  {\end{tabular}\\[\belowdisplayskip]}

\IEEEoverridecommandlockouts
\begin{document}

\title{Person Recognition using \\Smartphones' Accelerometer Data}


\author
{
\IEEEauthorblockN{Thingom Bishal Singha}
\IEEEauthorblockA{
NITK Surathkal\\
Email: bishalthingom@gmail.com}
\and
\IEEEauthorblockN{Rajsekhar Kumar Nath}
\IEEEauthorblockA{NITK Surathkal\\
Email: rajsekharkrnath@gmail.com}
\and
\IEEEauthorblockN{Dr. A. V. Narsimhadhan}
\IEEEauthorblockA{NITK Surathkal\\
Email: dhan257@gmail.com}
}

\maketitle

\begin{abstract}
Smartphones have become quite pervasive in various aspects of our daily lives. They have become important links to a host of important data and applications, which if compromised, can lead to disastrous results. Due to this, today\textquotesingle s smartphones are equipped with multiple layers of authentication modules. However, there still lies the need for a viable and unobtrusive layer of security which can perform the task of user authentication using resources which are cost-efficient and widely available on smartphones. In this work, we propose a method to recognize users using data from a phone’s embedded accelerometer sensors. Features encapsulating information from both time and frequency domains are extracted from walking data samples, and are used to build a Random Forest ensemble classification model. Based on the experimental results, the resultant model delivers an accuracy of 0.9679 and Area under Curve (AUC) of 0.9822.
\end{abstract}


\IEEEpeerreviewmaketitle

\section{Introduction}
User authentication and security of smartphones have
become issues of paramount importance as smartphones have become ubiquitous devices. With smartphones being one of the most important agents for the push towards digitization across the globe, there has been an ever increasing number of applications dealing
with financial transactions, health, contacts information,
etc. These applications generate an increasing amount of
critical information, the security of which is very much
essential and is also of great concern among users today \cite{HASP}. As a result, various methods of user authentication
in smartphones have surfaced over the past few years,
starting from the usual password based authentication to
pattern lock, fingerprint biometric authentication to even
face recognition \cite{apple}. However, all of these require active
user participation at the beginning and there is no way to
continuously authenticate the user at fixed intervals without
causing discomfort to the user.\\
In the past, certain methodologies and designs have been
proposed which attempt to use gait features from sensors
embedded into the smartphones to recognize user activity,
used for tasks such as activity monitoring, fall detection, and
so on. However, data from embedded sensors can also be
used to detect inherent patterns in the activities performed by
a particular user, which can be subsequently used to recognize
the person given the sensor data for activities carried out
by them \cite{btas10}. In this work, a system to identify the user
based on accelerometer data is presented. The system works
ubiquitously in the background, without needing the user to perform additional actions for authentication purposes.
For the task of user recognition, we train a Random Forest
ensemble classifier on a 31-features dataset extracted from
accelerometer data recorded during walking.\\
The presented work is organized as follows. Section II
introduces existing literature in gait analysis and use of
embedded sensors for activity and person recognition. The
methodology of the proposed work is described in section
III. The subsections describe the feature extraction process,
the Random Forest model and the validation method. This is
followed by section IV which presents the results achieved
by the model, followed by conclusion and future work in
section V.\\     
\section{Previous Work}
While delving into the sphere of person recognition using
activity data recorded using smartphones' accelerometers,
an important aspect is identifying the activity. In the past,
significant work has been done in this sphere, where activities
are identified using various supervised learning methods with
a great deal of accuracy. Any user authentication system
would require activity identification as one of the initial
layers. Once the activity has been identified, we can proceed
with the task of user recognition.
Some of the directly related works are of Johnston et al. \cite{btas10},  Lee et al. \cite{HASP} and Haong et al. \cite{gait}, where
works of similar nature have been attempted. In \cite{btas10}, a
strawman model using accelerometer data from smartphones
has been proposed for user identification. The strawman model
is then iterated over continuous 10-seconds samples over
a longer duration, following which the most voted person,
as identified by the model, is returned as the output. Time
domain features are used to generate a feature vector
for an identification window of 10 seconds. The dataset generated with
the corresponding feature vectors is used to train WEKA’s
J48 and Neural Network models. For the activity of walking,
an accuracy of 90.9\% was achieved by the Neural Network,
while the J48 model produced an accuracy of 84.0\%. On
the other hand, \cite{HASP} uses data from both smartwatches and
smartphones. In \cite{HASP}, the feature vector is composed of
both time domain (magnitude, mean, min, max, variance) and
frequency domain (amplitude of first peak, frequency and
amplitude of second peak of Discrete Fourier Transform) features with a focus on the integration of smartphone and
smartwatch, while considering identification from a number
of activities. The model used for classification/identification
is Kernel Ridge Regression. In a similar work \cite{gait}, SVM is used as the classifier. SVM and KRR are similar classifiers based on the kernel method. Kernel Ridge Regression (KRR)
\cite{svm} is a kernel method classifier model which uses the kernel trick. Instead of learning a fixed set of parameters for the input features, kernel classifiers instead learn a weight for each training example. The class prediction for the new inputs is carried out using a similarity function k, which is called the kernel, between the learned examples and the new input \cite{svm1}. Kernel based classifiers use a ”1 vs others” approach for multi-class classification problems, wherein the model is trained separately for each class separately \cite{svm}\cite{svm2}.
The basic premise of activity recognition and user
identification being that users perform activities differently,
or in other words they have a markedly different signature
for each activity. The work proposed in this paper combines
established powerful features used in activity recognition such
as magnitude, correlation, etc. with time domain features
used in past work on person recognition \cite{HASP}\cite{btas10}\cite{gait}
to selectively build a feature vector suitable for person the recognition task, while eliminating any redundant features, and at
the same time keeping the identification window minimal, i.e.,
just two seconds. A Random Forest ensemble classifier is used
which delivers a robust non-parametric model \cite{breiman}\cite{cart}.
\section{Methodology}

\subsection{Data Preprocessing}
\subsubsection{Data Collection}
~\\
Walking data is collected for 10 users using a smartphone which records tri-axial accelerometer readings at a frequency of 50Hz. For this, we use a Samsung Galaxy J-1 phone, which is kept in the side pocket of the users' trousers during data recording. The data is collected using OARS, a data collector application for Android smartphones.
\subsubsection{Feature Extraction}
~\\
The raw data is divided into identification intervals of 100 samples width with 50\% overlap.
The feature extraction is achieved by leveraging the methodologies of different activity recognition projects as discussed earlier. We picked up a basket of statistical features and added to it features like spectral centroid, widely used in audio recognition experiments \cite{speccentroid1}\cite{speccentroid2}. As done in many activity recognition problems \cite{featact1}\cite{featact2}, both time and frequency domain features are extracted. The Fast Fourier Transform (FFT) of the axial data of the identification windows for each of the axes is evaluated, which forms the base for all frequency domain features. 
The parameters extracted from the identification windows, which serve as the features for the classification problem are defined as follows:\\
(i) Mean: The mean values of the triaxial accelerometer data within each window are calculated for both the raw data and the FFT data for each of the three axes, which gives us a set of six mean values, abstracting data in both time and frequency domains.\\
(ii) Median: The median is calculated in a similar way as the mean. The median values are calculated for each of the axes taken in a similar way for both time and frequency domains for each of the three axes.\\
(iii) Magnitude: The magnitude is defined as the average of the root mean square of the tri-axial data (both time and frequency domain), and is calculated as follows:
\begin{equation}
Magnitude = (\sum_{k=1}^{l}\sqrt{x_{k}^2 + y_{k}^2 +z_{k}^2})/l\label{eq:1}
\end{equation}
\begin{conditions}
x_{k},y_{k},z_{k} & Instantaneous acceleration values\\
l & Length of window\\
\end{conditions}
The magnitude for frequency domain is calculated by putting instantaneous values of fourier transformed data in place of acceleration.\\
(iv) Cross-correlation: The cross-correlation is defined as the ratio of mean of x axis and z axis data and that of y axis and z axis data. The z axis is selected as the frame of reference as it remains constant for almost all possible orientations of the smartphone, and the ratios are taken with respect to z axis. The cross correlation of z axis with x axis and y axis are mathematically defined as follows:
\begin{equation}
Corr_{xz}=x_{mean}/z_{mean}\label{eq:2}
\end{equation}
\begin{equation}
Corr_{yz}=y_{mean}/z_{mean}\label{eq:3}
\end{equation}
\begin{conditions}

Corr_{xz}&Cross-correlation of x axis\& z-axis\\
Corr_{yz}&Cross-correlation of y axis\& z-axis\\
x_{mean}&Mean of acceleration values in x-axis\\
\end{conditions}
(v) Peak Count: Peak count for each axis refers to the number of local maxima for the axial data in the identification window. The average of the peak count over the three axes for the time domain data is selected as a feature.\\
(vi) Distance between Peaks: Distance between peaks refers to the average time interval between two successive peaks in a window. \\
(vi) Spectral Centroid: Spectral centroid is a measure used to characterise a spectrum. In this case, spectrum refers to the identification window of acceleration values. It indicates where the center of mass of the spectrum is\cite{wiki}. The spectral centroid of each window for the three axes using the FFT values as weights is given by:
\begin{equation}
Centroid=\sum_{k=1}^{l} x_{tk}*f_{tk}/l \label{eq:4}
\end{equation}
\begin{conditions}
x_{tk} & Instantaneous acceleration\\
f_{tk} & Instantaneous value of FFT\\
l & length of window\\
\end{conditions}
(vii) Average Difference from Mean: The absolute difference from mean of the window for each axis (time domain) is calculated as follows: 
\begin{equation}
Diff_{x}=Avg(|x_{t}-x{mean}|)
\end{equation}
\begin{conditions}
Diff_{x} & Difference from mean\\
x_{t} & Instantaneous acceleration\\
x_{mean} & Mean as defined in eq. \ref{eq:1}\\
\end{conditions}
The plots of comparison of two of the users for magnitude, cross-correlation(xz) in frequency domain, and cross-correlation(yz) in frequency domain are shown in figures \ref{pic1},\ref{pic2} and \ref{pic3}, respectively. The features were extracted using numpy library and the resultant feature vectors were then labeled with the respective users.
\begin{center}
\begin{figure}[!ht]
\includegraphics[scale=0.4]{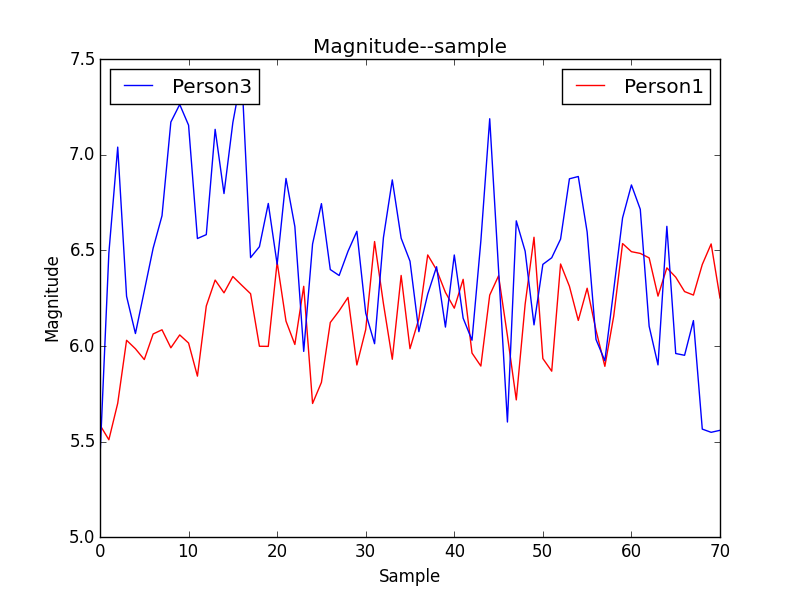}
\caption{Magnitude: Person 1 vs Person 3}
\label{pic1}
\end{figure}
\begin{figure}[!ht]
\includegraphics[scale=0.4]{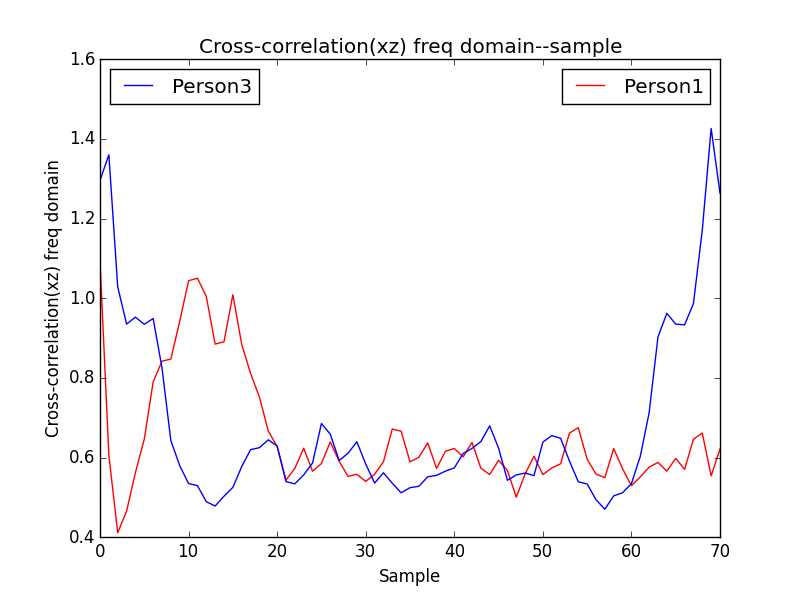}
\caption{Cross-correlation(xz) in Freq Domain: Person 1 vs  Person 3}
\label{pic2}
\end{figure}
\begin{figure}[!ht]
\includegraphics[scale=0.4]{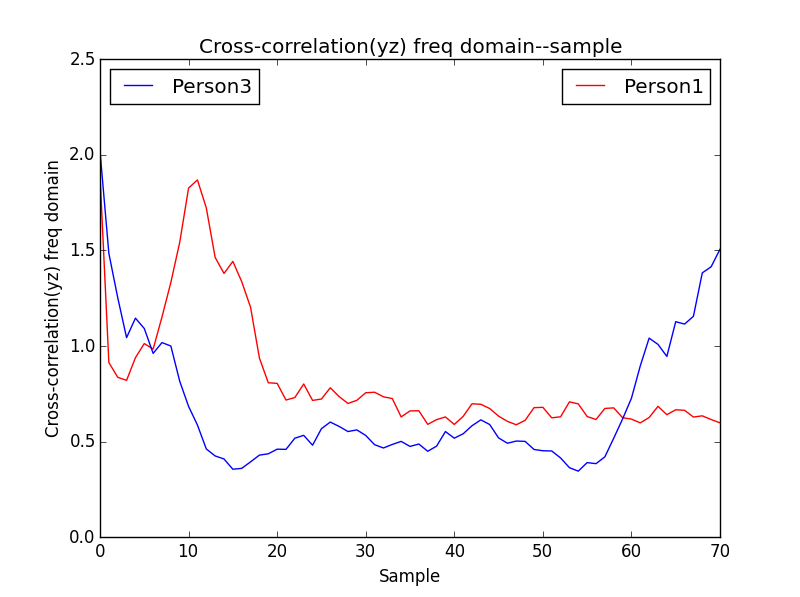}
\caption{Cross-correlation(yz) in Freq Domain: Person 1 vs  Person 3}
\label{pic3}
\end{figure}

\end{center}
\subsubsection{Dataset Creation}
~\\
Using features extracted from the raw accelerometer data as described in the previous section, a dataset is generated where each row corresponds to an interval of 100 continuous samples, with the output label being the person the samples are from. Evaluation of any learning model involves splitting of the dataset into a training set and a testing set to check how the model performs on a dataset it hasn't seen. However, this method is susceptible to high variance. The evaluation may depend heavily on which data points end up in the training set and which end up in the test set, and hence, the performance of the model may significantly vary depending on the division into training and testing datasets. This is overcome by k-fold cross-validation in which the original dataset is split into k equally sized subsets. Of these k subsets, k-1 subsets are used for training while the remaining one subset is used for testing the model. This process is repeated k times so that each observation is used for validation only once. The average of the k results is calculated to arrive at a single estimation for the model \cite{cmucv} \cite{skcv}. In this work, we use a stratified 10-fold classification model so that each fold (subset) is representative of the whole dataset. 

\subsection{Model}
We use a Random Forest Classifier as our classification model, which is an ensemble of 64 decision trees generated from a set of 30 features. Random Forest (RF) Classifier is a learning-based classification algorithm which relies on an ensemble of multiple decision tree classifiers. Taking advantage of two powerful machine-learning techniques in bagging and random feature selection, the RF classifier combines the output of individual decision trees which are generated by selecting a random subset of the features. 
\subsubsection{Decision Tree Classifier}
~\\
The model used in this work is implemented with scikit-learn\textquotesingle s Random Forest Classifier which uses an optimized version of Classification and Regression Trees (CART) classification algorithm, proposed by  Breiman et  al. \cite{breiman}, for building the decision trees. CART is a non-parametric learning algorithm which generates a classification or regression tree. Each decision node of the tree splits the data into groups, with the attribute being chosen such that the resultant groups are increasingly homogenous as we move down the tree \cite{cart}. \\
In classification problems, CART uses the Gini impurity measure for producing homogenous groups.The Gini impurity measure at a node for a category k is defined as: 

\begin{equation}
G(p_{k}) = p_{k}*(1-p_{k})
\end{equation}
\begin{conditions}
 p_{k}     &  Proportion of observations in class k
\end{conditions}

Impurity at a node n is defined as the sum of impurities for all categories \cite{gini}, and is given by:
\begin{equation}
I_{n} = \sum_{k=1}^{K}G(p_{k})
\end{equation}
\begin{conditions}
 I_{n}     &  Impurity at node n\\
 G(p_{k})     &  Gini impurity for class k
\end{conditions}
The CART algorithm considers all possible splits across the input features and selects the feature which maximizes the drop in impurity \cite{gini}, defined as:
\begin{equation}
\Delta I = p(n_{0})I(n_{0})-(p(n_{1})I(n_{1})+p(n_{2})I(n_{2}))
\end{equation}
\begin{conditions}
 \Delta I     &  Change in impurity\\
 n_{0}     &  Parent node\\   
 n_{1}, n_{2} &  Children nodes \\
 p(n) & Ratio of observations at the node n
\end{conditions}

\subsubsection{Random Forest Ensemble}
~\\
Since Decision Trees are non-parametric classifiers, they are suitable for datasets which are not linearly separable. They are also robust to classifiers due to the splitting during the tree generation. However, Decision Trees are often associated with high variance since the branches made by the splits are enforced at all lower levels of the tree. As a result, a slight error in the data could result in a considerably different sequence of splits, resulting in a different classification rule \cite{opitz}. A bootstrap is a random sub-sample of the dataset. In Random Forests, multiple Decision Trees are generated using bootstrap samples and the result is determined by aggregating the output of the individual trees. While selecting the feature for splitting at a node, the RF algorithm only uses a random subset of features. This reduces the correlation between the individual trees. The architecture of the RF Classifier is shown in figure \ref{rfpic} \cite{rfimage}:

\begin{figure}[!ht]
\begin{center}
\includegraphics[scale=0.3]{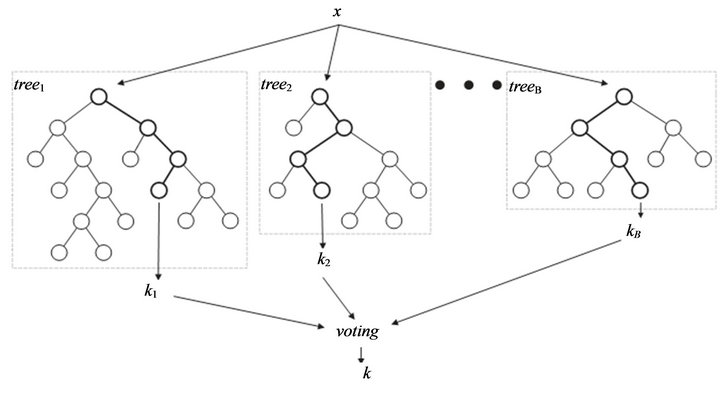}
\end{center}
\caption{Architecture of Random Forest Classifier}
\label{rfpic}
\end{figure}

\section{Results and Analysis}
The dataset for our model was prepared by collecting raw accelerometer data and then extracting features from this time-series data using the process described in section III A. The resultant dataset of 31 input features consisted of ~ 3600 examples.
\subsection{Metrics}
In this section, we define the metrics used to measure the performance of the classifier.\\
For a binary classification problem, its performance can be determined by computing the number of correctly labeled positive observations (true positives), the number of correctly labeled negative observations (true negatives), the number of negative observations incorrectly labeled as positive (false positives) and the number of positive observations incorrectly labeled as negative (false negatives) \cite{performancemetrics}. These four values constitute a confusion matrix as shown in Table 1.

\newcommand\MyBox[2]{
  \fbox{\lower0.75cm
    \vbox to 1.7cm{\vfil
      \hbox to 1.7cm{\hfil\parbox{1.4cm}{#1\\#2}\hfil}
      \vfil}%
  }%
}

\noindent
\renewcommand\arraystretch{1.5}
\setlength\tabcolsep{0pt}

\begin{tabular}[!ht]
{c >{\bfseries}r @{\hspace{0.7em}}c @{\hspace{0.4em}}c @{\hspace{0.7em}}l}

  \multirow{10}{*}{\parbox{1.1cm}{\bfseries\raggedleft Actual\\ Value}} & 
    & \multicolumn{2}{c}{\bfseries As Classified by Model} & \\
  & & \bfseries Positive$'$ & \bfseries Negative$'$ \\
  & Positive & \MyBox{True}{Positive} & \MyBox{False}{Negative} \\[2.4em]
  & Negative & \MyBox{False}{Positive} & \MyBox{True}{Negative} \\
\end{tabular}
\begin{center}
TABLE 1: Confusion Matrix
\end{center}

Recall, Specificity and Area under Curve (AUC) are metrics originally used for binary classification problems. However, in our case, we have multiple classes (equivalent to the number of users in dataset). Due to this, for each class, a "1 vs others" approach is used \cite{performancemetrics}. \\
Recall, Specificity and Area under Curve (AUC) are calculated for each of the classes with the class in consideration being the positive class, while the other classes are interpreted as negative. The final values are generated by calculating a weighted average (micro-averaging) of the respective values for each class, where weight of each class is given by:
\begin{equation}
W_{c} = n_{c}/n
\end{equation} 
\begin{conditions}
 W_{c}     &  Weight of class c\\
 n_{c}     &  Number of actual observations in class c\\   
 n &  Total number of observations
\end{conditions}

The metrics used for evaluating the model are defined as follows:
\subsubsection{Accuracy}
Accuracy is the ratio of correctly predicted observations to the total observations and indicates the overall effectiveness of a classifier and is given by:
\begin{equation}
Accuracy = n_{Correct}/n
\end{equation} 
\begin{conditions}
 n_{Correct} & No. of correctly labeled observations\\
 n     &  Total number of observations
\end{conditions}

\subsubsection{Recall (Sensitivity)}
Recall is the ratio of correctly labeled positive observations to the total number of actual positive observations and indicates the effectiveness of classifier in identifying positive observations \cite{performancemetrics}, and is defined as:
\begin{equation}
Recall = W_{c}*R_{c}
\end{equation}
\begin{equation}
R_{c} = (tp_{c})/(tp_{c} + fn_{c})
\end{equation} 
\begin{conditions}
 R_{c}     &  Recall of class c\\
 tp_{c}     &  True positives in class c\\
 fn_{c}     &  False negatives in class c
\end{conditions}

\subsubsection{Specificity}
Specificity is the ratio of correctly labeled negative observations to the total number of actual negative observations and indicates the effectiveness of classifier in identifying negative observations \cite{performancemetrics}, and is defined as:
\begin{equation}
Specificity = W_{c}*S_{c}
\end{equation}
\begin{equation}
S_{c} = (tn_{c})/(fp_{c} + tn_{c})
\end{equation} 
\begin{conditions}
 S_{c}     &  Specificity of class c\\
 tn_{c}     &  True negatives in class c\\
 fp_{c}     &  False positives in class c
\end{conditions}

\subsubsection{Area under Curve (AUC)}
Area under Curve (AUC) is the macro-average of Recall and Specificity and indicates the classifier's ability to avoid false classification \cite{performancemetrics}, and is defined as:
\begin{equation}
AUC = W_{c}*AUC_{c}
\end{equation}
\begin{equation}
AUC_{c} = (R_{c} + S_{c})/2
\end{equation} 
\begin{conditions}
 AUC_{c}     &  Area under Curve for class c
\end{conditions}

\subsection{Number of Trees in the Forest}
\begin{figure}[!ht]
\includegraphics[scale=0.55]{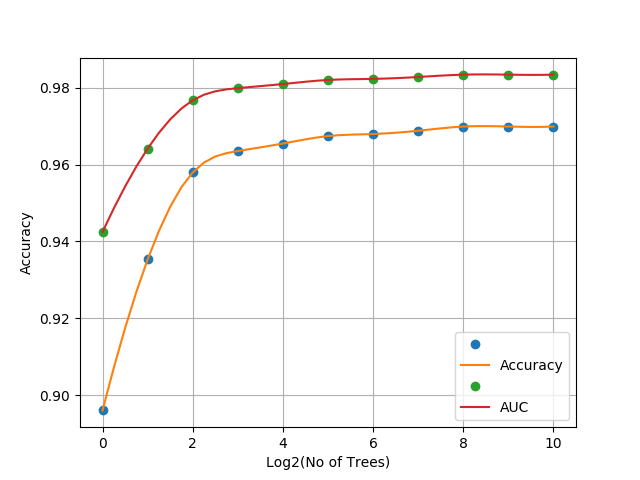}
\caption{Effect of number of trees on performance of RF Classifier}
\label{rftrees}
\end{figure}

The time complexity of a Random Forest is given by:
\begin{equation}
O( t * k_{try} * nlog(n) )
\end{equation} 
\begin{conditions}
 t     &  Number of trees in ensemble\\
 k_{try} & Number of features\\
 n & Number of records
\end{conditions}
In general, more the number of trees, better the performance of the model. But beyond a point the performance starts to plateau and the additional time overhead isn't commensurate with the performance improvement. For our model, the performance, both in terms of accuracy, as well as area under curve, plateaus as the number of trees reaches 64 as shown in figure \ref{rftrees}.

\subsection{Person Recognition Results}
In this section, the results for Person Recognition using the RF Classifier model are presented. In addition to the RF classifier, we also present the results obtained by using three other popular classification models using the same set of features. The models used in addition to the RF classifier are
Logistic Regression (LR), Support Vector Machine (SVM), and Decision Trees (DT) (defined in section III).
\subsubsection{Logistic Regression}
Logistic Regression (LR) is a classification model which uses linear regression and the logit function to predict the category of the output variable. The value generated by linear regression is passed through the logit function which maps the value between 0 and 1 and represents the probability of the input vector belonging to category k \cite{logit}.
\subsubsection{Support Vector machine}
Support Vector Machine (SVM) is a discriminative classifier and is defined by a hyperplane. The hyperplane refers to an (n-1) dimensional plane which gives the optimal separation of the training examples into two output classes. For a multiclass problem, the SVM is run for all classes using a ”1 vs others” approach \cite{svm2}

The performance of the models over accuracy, AUC and
recall is shown in table 2 and the person-wise accuracy is
shown in table 3.\\

\begin{center}
\begin{tabular}{ |P{2cm}|P{2cm}|P{2cm}|P{2cm}|  }
 \hline
 Model/Metric& Accuracy& AUC& Recall \\
 \hline
 RF   & 0.9679 & 0.9823 & 0.9966\\
 DT & 0.9613 & 0.9784 & 0.9955\\
 LR    & 0.8768 & 0.9314 & 0.9861\\
 SVM  &0.7158 &0.8422 & 0.9685\\
 \hline
\end{tabular}
\end{center}

\begin{center}
TABLE 2: Performance of Classification Models
\end{center}
~\\
\begin{center}
\begin{tabular}{ |P{1.5cm}||P{1.5cm}|P{1.5cm}|P{1.5cm}|P{1.5cm}|  }
 \hline
 Person& RF & DT & LR & SVM \\
 \hline
1&0.9443&0.9619&0.9032&0.8592   \\
2&0.9623&0.9623&0.8899&0.5797   \\
3&0.9683&0.9405&0.8651&0.4524   \\
4&0.9615&0.9586&0.7278&0.6361   \\
5&0.9385&0.9198&0.7273& 0.6872  \\
6&0.997& 0.9881&0.9941&0.9555   \\
7&0.9882&0.9794&0.7971& 0.7441  \\
8&0.9853&0.9765&0.868 & 0.3196 \\
9&0.9692&0.9513&0.9538& 0.7974  \\
10&0.968&0.9699&0.9812&0.9303   \\

 \hline
\end{tabular}
\end{center}
\begin{center}
TABLE 3: Person-wise Accuracy
\end{center}

The Random Forest (RF) classifier outperforms the Decision
Tree (DT) classifier due to advantages of bagging and randomized
feature selection as described in section III B. The Logistic
Regression model is stumped by both RF classifier and the DT
classifier. This stems from the limitations of LR classifiers in
handling large feature vectors and large number of categorical
outputs. The Support Vector Machine (SVM) classifier is also trumped by the RF and
DT classifier. This stems from SVMs being originally designed
for binary classification problems. Additionally, SVMs also
perform poorly for imbalanced classes.
In related work, the model proposed by Johnston et al.
\cite{btas10} produced an accuracy of 90.9\% using a Neural Network
model and an accuracy of 84.0\% using WEKA's J48 model
on a dataset containing 2081 samples. On the other hand, the model proposed by \cite{HASP} delivered an accuracy of 92.1\% using kernel Ridge Regression on a dataset containing 800 samples. It must be noted that for the use case of authentication,
an important measure would be the model\textquotesingle s ability to avoid
misclassification as it is essential to make sure the right person
is identified, i.e, false negatives are minimized to avoid an
inconvenient user experience, and false positives are minimized
to avoid false authentication. This is well represented by the
Area under Curve (AUC) metric as described previously.\\

\section{Conclusion and Future Work}
In this paper, we proposed a system using the Random Forest classification model for person recognition using smartphone's accelerometer data. The feature set for the dataset was extracted from the time-series accelerometer data in both time domain, and frequency domain. The ubiquitous nature of smartphones and the activity of walking, along with the numerical nature of accelerometer data, makes this method of authentication unobtrusive and computationally cheaper as compared to vision-based recognition models. The experimental results achieved by our model demonstrate the potential of integrating accelerometer based person recognition with biometric authentication.\\
In future, we aim to further improve the AUC of the model due to the detrimental effects misclassification could have for the proposed authentication use case. We also plan to increase the number of users and incorporate more natural activities and more positions for keeping the smartphone during data collection, such as back pocket, shirt pocket, etc.


%

\end{document}